\documentclass[useAMS,usenatbib]{mn2e}
\usepackage{graphicx}
\usepackage{booktabs}
\usepackage{color}

\newcommand{\apj}{ApJ}
\newcommand{\apjl}{ApJL}
\newcommand{\apjs}{ApJS}
\newcommand{\aj}{AJ}
\newcommand{\mnras}{MNRAS}
\newcommand{\nat}{Nature}

\newcommand{\araa}{ARA\&A}
\newcommand{\aap}{A\&A}

\definecolor{darkgreen}{rgb}{0.0,0.5,0.0}
\definecolor{darkred}{rgb}{0.5,0.0,0.0}
\definecolor{brown}{rgb}{0.65,.16,0.16}
\definecolor{grey}{rgb}{0.4,0.5,0.6}

\title[Impacts of accretion and outflows on the SFR]{Feedback by supermassive black holes in galaxy evolution: impacts of accretion and outflows on the star formation rate}

\author[M. Raouf et al.]{Mojtaba Raouf $^{1,2}$\thanks{E-mail:
		m.raouf@ipm.ir},
	    Joseph Silk $^{3,4,5}$,
		Stanislav S. Shabala $^{6,7}$,
		Gary A. Mamon $^{3}$,
		\newauthor
		Darren J. Croton $^{8}$,
		Habib G. Khosroshahi $^{1,3}$,
		Ricarda S. Beckmann$^{3}$
	\\	
	$^{1}$School of Astronomy, Institute for Research in Fundamental Sciences (IPM), Tehran, 19395-5531, Iran\\
	$^{2}$Korea Astronomy and Space Science Institute, 776 Daedeokdae-ro, Yuseong-gu, Daejeon 305-348, Korea\\
	$^{3}$Institut d'Astrophysique de Paris (UMR 7095: CNRS \& Sorbonne Universit\'e), 98 bis Bd. Arago, F-75014 Paris, France\\
	$^{4}$Department of Physics and Astronomy, Johns Hopkins University, Baltimore, MD 21218, USA\\
	$^{5}$BIPAC, Department of Physics, University of Oxford, Keble Road, Oxford OX1 3RH, UK\\
    $^{6}$School of Natural Sciences, Private Bag 37, University of Tasmania, Hobart, TAS 7001, Australia\\
    $^{7}$Centre for Astrophysics Research, School of Physics, Astronomy and Mathematics, University of Hertfordshire, College Lane, Hatfield,\\ Hertfordshire AL10 9AB, UK\\
    $^{8}$ Centre for Astrophysics \& Supercomputing, Swinburne University of Technology, PO Box 218, Hawthorn, Victoria 3122, Australia.
}

\begin{document}
	
	\date{}
	\pagerange{\pageref{firstpage}--\pageref{lastpage}} \pubyear{2012}
	\maketitle
	\label{firstpage}

	\begin{abstract}
		
We describe a physical model of the outflows produced as a result of gas accretion onto a black hole, and the resultant changes to star formation rates and efficiencies in galaxies, using the \textsc{Radio-SAGE} semi-analytic galaxy formation model. 
We show that the ratio of outflow rate to SFR of galaxies is mainly driven by black hole mass and virial halo mass, and show that the SFR is higher than the outflow rate at low black hole masses. The model consistently reproduces the observed evolution of star formation rate density from $z = 6$ to $z = 0$, as well as the trend of the stellar mass - halo mass relations.  We show the characteristic growth of massive galaxies influenced by AGN feedback at different redshifts. We find feedback to be prevalent in the most massive galaxy halos, inhibiting the cooling catastrophe. 

	\end{abstract}


	\section{Introduction} 
	
	One of the major goals of modern astronomy is to understand the star formation history of the Universe.  At the present epoch, the cosmic star formation rate (SFR) density increases to $z\simeq 1$ \citep[e.g. ][]{Hopkins2006, Villar2008, Sobral2009} then reaches a peak around $z\simeq 2$ and decreases at higher redshifts \citep{PerezGonzalez2008, Bouwens2008}.
	There have been extensive efforts to understand the evolution of galaxy SFRs through the relatively tight correlation between galaxy SFR and stellar mass known as the galaxy SFR main sequence, with a shallower slope at lower redshift \citep{Elbaz2007, Daddi2007, Salim2007, Oliver2010, Daddi2009a, Stark2009, Gonzalez2009, Pearson2018}. 
	Furthermore, the star formation main sequence galaxies show much longer typical depletion times with respect to the entire mass range of star forming galaxies via studies of the dust and gas content of $z$$\sim$2 galaxies  \citep{magdis12b,bethermin15,genzel15,tacconi18}.	
		
	Every massive galaxy could host a supermassive black hole (SMBH) at the center. This provides a mechanism for explaining its presence of Active Galactic Nuclei (AGN), which suppress excessive star formation and hence galaxy growth \citep{Silk1998}.
	A principal motivation to include AGN feedback in galaxy formation models is to understand how the suppression of rapid gas cooling then leads to a suppression of the formation of new stars. In the most massive haloes, the ratio of galaxy stellar mass to dark matter halo mass decreases with increasing mass \citep{Behroozi2013a}, rendering supernova feedback inefficient \citep{Dubois2008,Dashyan2018}. AGN feedback could provide an effective method for quenching massive galaxies, as well as regulating the growth of supermassive black holes (SMBH) \citep{Binney1995,Benson2003,DiMatteo2005,Bower2006,Croton2006,Sijacki2007,Cattaneo2009,Fabian2012}.
	
	The gas content of the galaxy could directly be influenced by feedback from the AGN jets, which heat up the surrounding gas leading to negative feedback on star formation\footnote{AGN feedback might even accelerate star formation by further compressing the cold gas of the galaxy, in a so-called positive feedback mode \citep{Silk2005, Crockett2012, Gaibler2012, Santini2012, Bieri2015}}. At higher redshift, the  powerful energy-driven winds from quasars could lead to a rapid decline of the  star formation rate \citep{Maiolino2012,Page2012,Farrah2012,Cicone2014,Costa2015,Williams2017}. Observationally, there are correlations between the properties of  supermassive black holes and host galaxies in the different modes of accretion power \citep{Ferrarese2000,Gebhardt2000,DiMatteo2008,Booth2009,Dubois2012,Sijacki2015,Volonteri2016}. Observational evidence shows that 
there are significant outflows in massive galaxies using only \(5-10\% \) of accretion power \citep{Moe2009,Saez2009,Dunn2010} including highly uncertain measurements with many assumptions going into efficiency estimates, and in part due to different AGN phenomena other than the assumptions of the jet model.
By using a larger fraction of accretion power, the \textsc{Radio-SAGE} model provides the necessary energy input to both quench star formation and to avoid the overcooling problem, and reproduces both the observed optical and radio luminosity functions at the present epoch \citep{Raouf2017}.

	Black hole growth strongly depends on its environment.  For instance, \cite{Raouf2016} using hydrodynamical simulations show that the black holes hosted by the brightest group galaxies in dynamically young groups are more massive than those with a similar stellar mass but residing in dynamically old groups \citep[for a definition of old and young groups see][]{Raouf2014} resulting in a  strong dependency of the black hole growth on its environment. They also showed that such old groups with massive black holes have a lower rate of black hole accretion in comparison to the young systems, in agreement with observations and semi-analytic model predictions for radio luminosity  \citep{Khosroshahi2017, Raouf2018}.

	The underlying physics of black hole accretion, and especially the mechanisms connecting the accretion flow with large-scale outflows, are still outstanding problems in astrophysics. 
	Here, we address the following questions, \textit{How} does accretion of the SMBH affect the SFR? and \textit{what} is the connection between SFR and AGN outflow rate in the galaxies hosted by a massive black hole? These are debated in a number of publications \citep[e.g. ,][]{Quintero2004, Thomas2005, Schawinski2014, Kaviraj2015, Shabala2017}.
	The rest of the paper will focus on the relationship between supermassive black hole growth and outflows. This paper is organized as follows: In Section 2, we describe our N-body and \textsc{Radio-SAGE} model framework with publicly available code as a sub-group of the original SAGE repository\footnote{https://github.com/mojtabaraouf/sage}. In Section 3 we describe the constraints on our model. Results and predictions are discussed in Section 4. We present the summary of our results in Section 5.

	\section{The galaxy formation model; Radio-SAGE} \label{ModelDescribtion}
	We only give a brief introduction to the base \textsc{Radio-SAGE} galaxy formation model here and refer the interested reader to \cite{Raouf2017} and the original \textsc{SAGE} paper \citep{Croton2016} for a full description. We use the Millennium Simulation \citep{Springel2005} N-body dark matter halo merger trees as input into \textsc{Radio-SAGE}, which has been updated with a new gas density profile and AGN physics calibrated to match key observations, as explained below.
	
The Millennium Simulation, containing $2160^3$ particles of mass $8.6 \times 10^{8}\, h^{-1} {\rm M_{\odot}}$ within the box volume of $(500\, h^{-1}\rm{Mpc})^3$, was run using the popular \textsc{GADGET-2} code and adopted a cosmological model consistent with the first year Wilkinson Microwave Anisotropy Probe data \citep[WMAP-1,  with parameters $\Omega_{\rm m} = 0.25$, $\Omega_{\Lambda} = 0.75$ and $H_0 = 100\,h\,\rm km\,s^{-1}\,Mpc^{-1}$ where $h=0.73$]{Spergel2003}. The simulation covers redshift $z$=127 to the present epoch and stores its data in 64
separate snapshots. The SUBFIND algorithm \citep{Springel2001} was then applied to the Friends-of-Friends \citep[FOF;][]{Davis1985}  catalog to identify subhalos by restricting the boundaries of substructures. The halo mass resolution is 20 particles, while halo merger trees are provided by the \textsc{L-HALOTREE}. 

Following our previous study to ensure \textsc{Radio-SAGE} produces a galaxy population akin to that observed around us, it is calibrated to statistically match a set of key observables up to $z = 7$, with a main focus on star formation. It uses a more realistic hydrostatic hot gas density profile to calculate cooling, the AGN jet can heat the hot gas to higher temperatures than the original \textsc{SAGE}, and the cooling rate is affected by both this new temperature and the jet inflated cavities. 
	
\subsection{Cooling and black hole accretion}
In this model, baryons \footnote{Here, the mass fraction in baryons associated with every dark matter halo is taken to be $f_b = 0.17$, consistent with the WMAP1 results of \cite{Spergel2003}.} initially form as diffuse hot gas with primordial composition around the galaxy. We adopted the density profile described by \cite{Makino1998}, which closely matches the observationally fit $\beta$-model but with a robust theoretical foundation \citep{Yates2018}. 

Supermassive black holes grow by merging and accretion of gas during major mergers of galaxies as same as the original SAGE model \citep{Croton2006}. The black hole accretion adopted by the ratio of accreted mass to total available cold gas mass scales with halo virial velocity \citep{Kauffmann2000} can be written as
	\begin{equation}
	M_{\rm acc} = \frac{ f_{\rm BH}\  m_{\rm R}\ M_{\rm Cold} } {1 + (280 ~\rm km\,s^{-1}/ V_{\rm vir})^2}, 
	\end{equation}	
	where $m_{\rm R}$ = $m_1 / m_2$ is the mass ratio of merging galaxies (for major merger $m_{\rm R}\!>\!0.3$), and $f_{\rm BH} = 0.015$ is the black hole growth rate. Major mergers are sufficiently energetic that the disk of the central galaxy is also destroyed and its stars added to the bulge which grows from the stellar remnants of merged satellites. Moving beyond this original accretion implementation, we built a physical model of the outflows produced as a result of gas accretion onto a black hole. The accretion rate of gas feeding the black hole (following description in  \cite{Croton2006})  is approximated by the Bondi-Hoyle formula \citep{Bondi1952}, ($\dot{m}_{\rm BH}\equiv \dot{M}_{\rm BH}/\dot{M}_{\rm edd}$)
	\begin{equation}\label{Bondi}
	\dot m_{\rm BH} = \frac{2.5 \pi G^2 m_{\rm BH}^2 \rho_0}{c_s^3} ~,  
	\label{eq:BHaccrOrig}
	\end{equation}
	where $m_\mathrm{BH}$ is the black hole mass and $\rho_0$ is the density of accreting hot gas around the black hole. $c_s \equiv V_{\rm vir}$ and $G$ are the speed of sound in the gas and the gravitational constant, respectively. Recently it has been argued \citep{Hardcastle2018} that chaotic cold accretion \citep{Gaspari2013} may provide a better description than Bondi accretion; we defer this issue to future work.  
	\begin{figure*}
	\centering
	\includegraphics[width=0.95\textwidth]{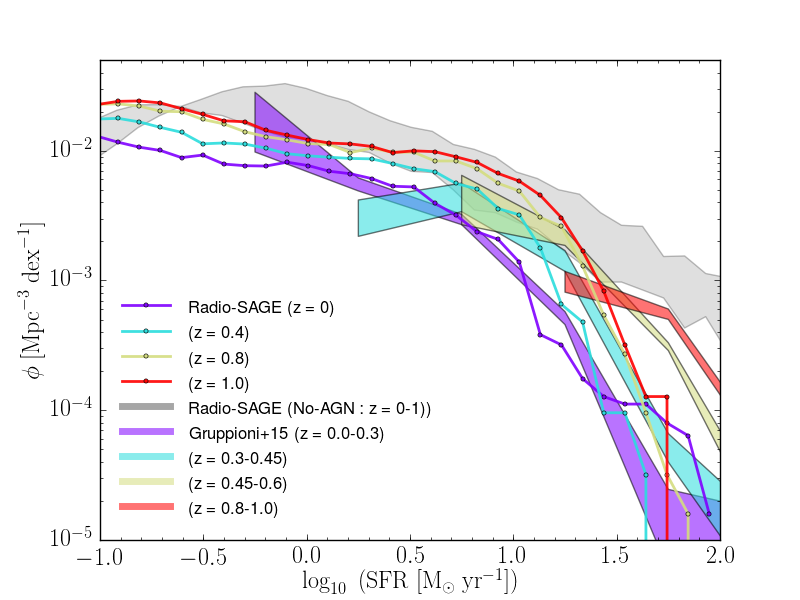}
	\caption{Distribution of galaxy star formation rates in different redshift bins from $z = 0$ to 1. Solid lines show Radio-SAGE predictions. Shaded curves are observations by \citet{Gruppioni2015} at $0<z< 1$. The grey shaded region shows the model without AGN feedback for the full redshift range .}
	\label{fig:SFRF}
\end{figure*} 

\begin{figure*}
	\centering
	\includegraphics[width=.94\textwidth]{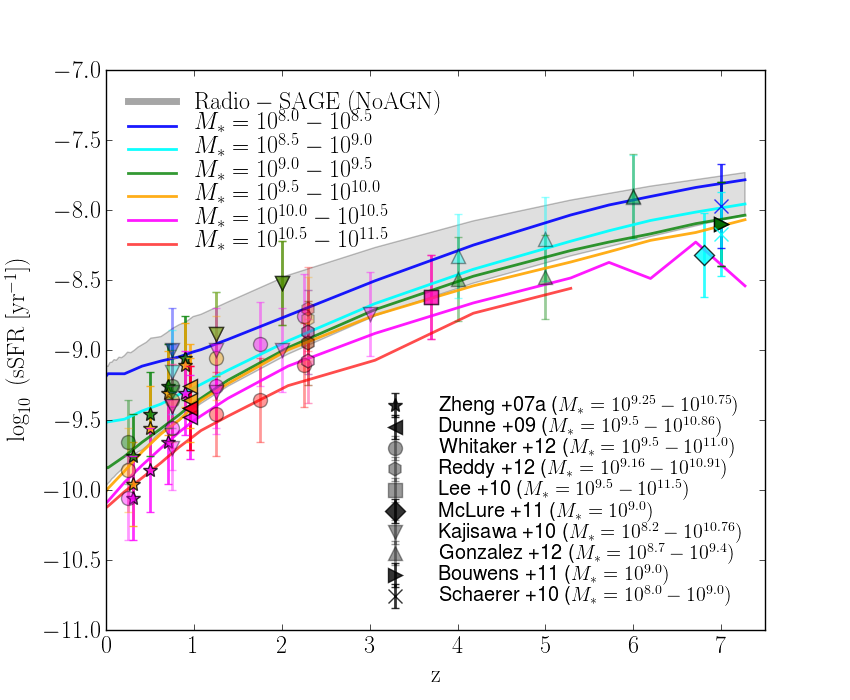}
	\caption{Redshift evolution of the median specific star formation rate from $z = 0$ to 7,  for different stellar mass bins. The observed evolution is indicated by symbols, where the colour of each symbol denotes the relevant stellar mass range. The vertical error bars indicate the estimated intrinsic scatter in sSFR. The grey shaded region shows the model without AGN feedback for all masses from \textsc{Radio-SAGE} .  
	}
	\label{fig:SFR_Z}
\end{figure*} 

\begin{figure*}
	\centering
	\includegraphics[width=0.497\textwidth]{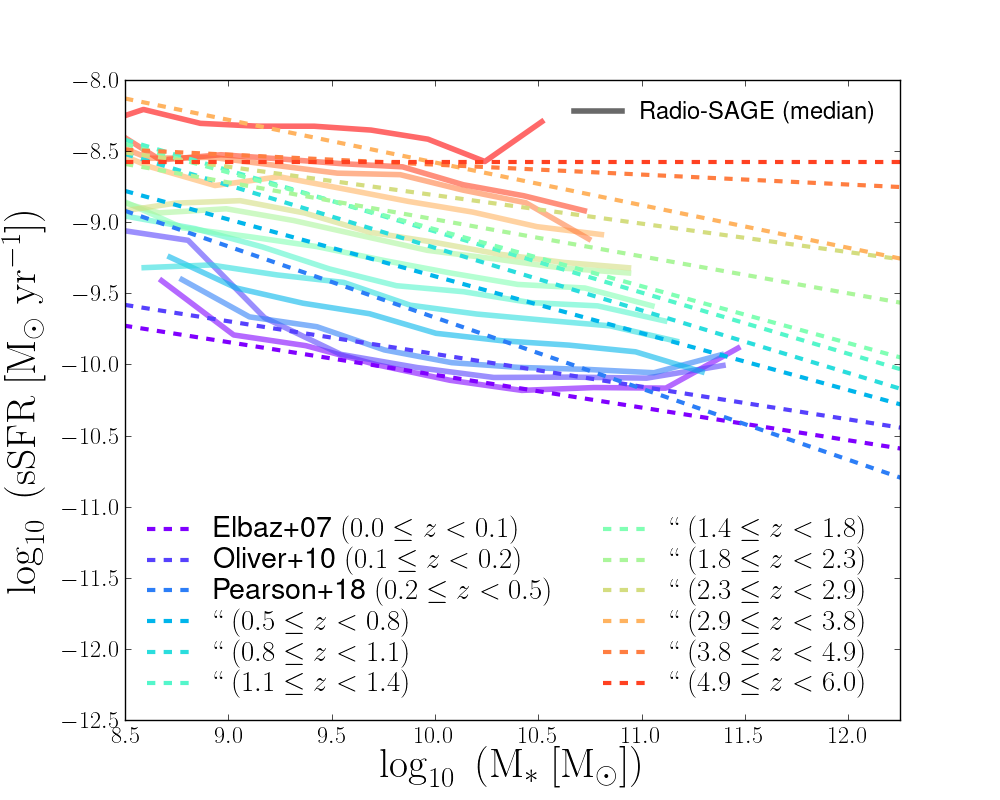}
	\includegraphics[width=0.497\textwidth]{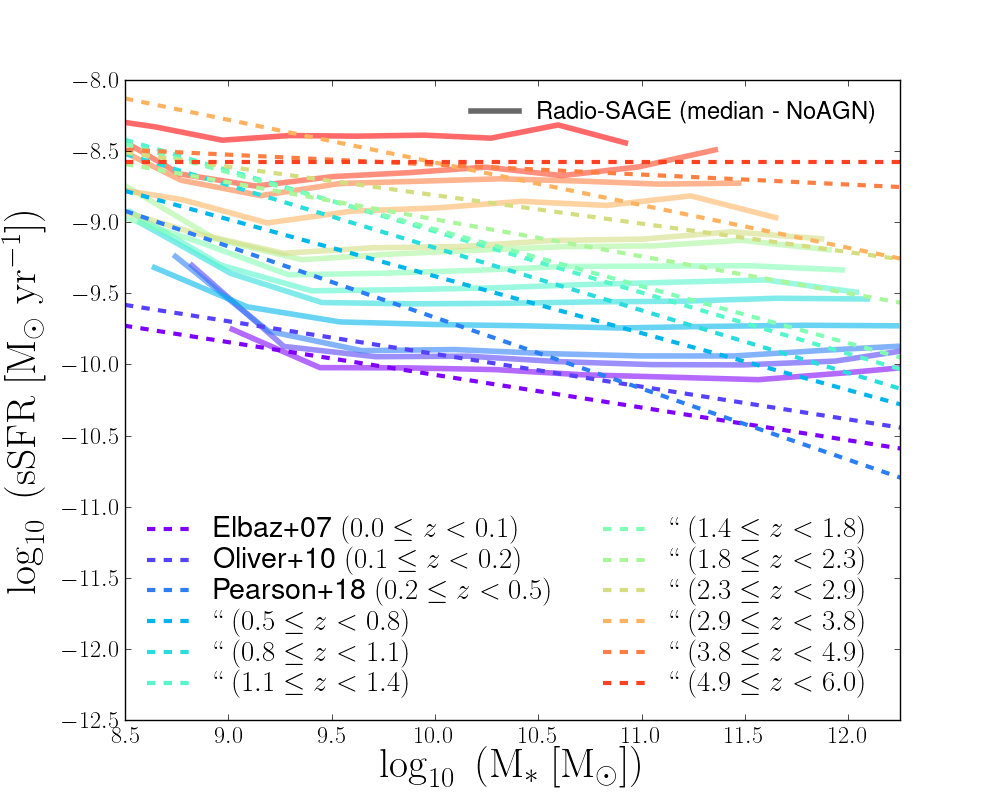}
	\caption{Left: Evolution of the sSFR - stellar mass relation. Solid lines are median binned data to the \textsc{Radio-SAGE} main sequence. Dashed lines are observed trends. Right: Same but for \textsc{Radio-SAGE} without AGN feedback .}
	\label{fig:SFR_MS}
\end{figure*} 
	
	\begin{figure}
	\centering
	\includegraphics[width=0.5\textwidth]{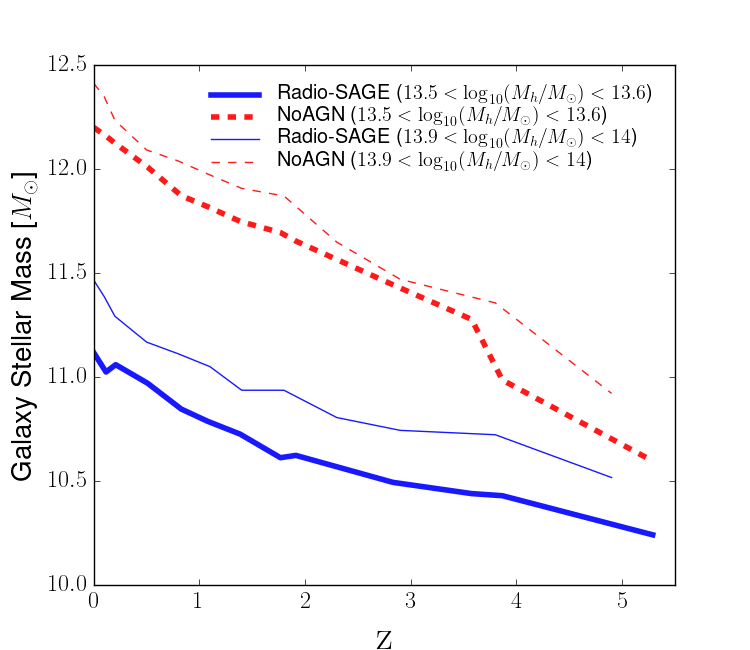}
	\caption{ The median stellar mass at each redshift, for two different final halo masses in the model with (solid-line) and without (dashed-line) AGN.}
	\label{fig:AGN_Scale}
    \end{figure} 

	\begin{figure*}
		\centering
		\includegraphics[width=0.45\textwidth]{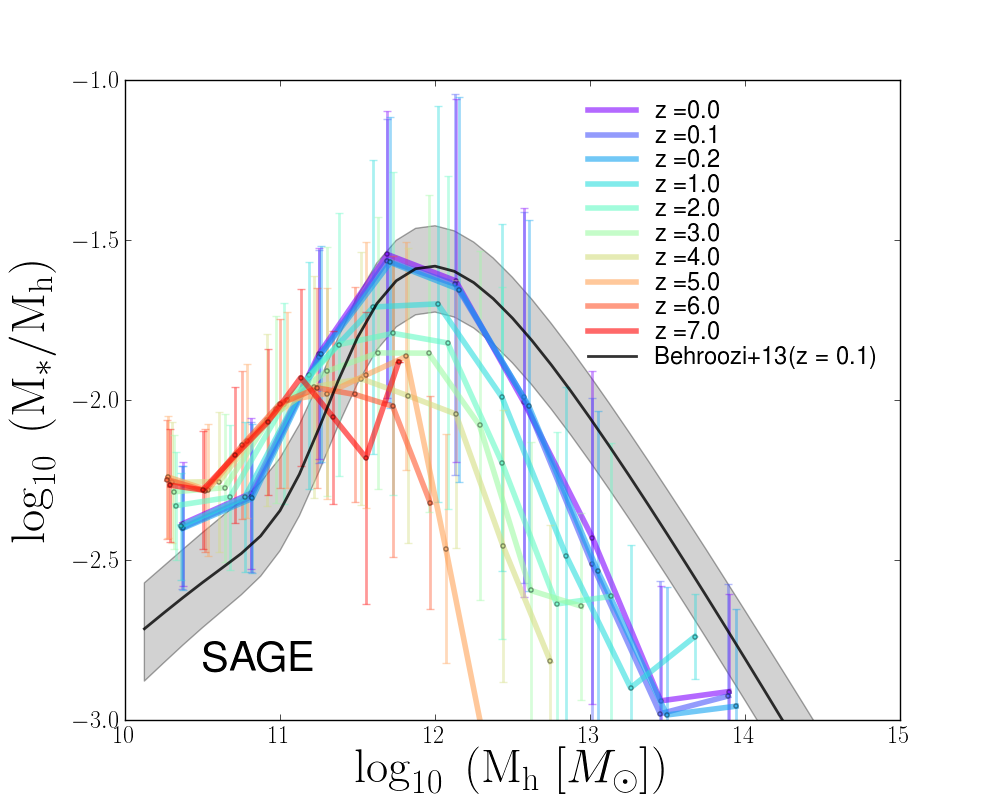}
		\includegraphics[width=0.45\textwidth]{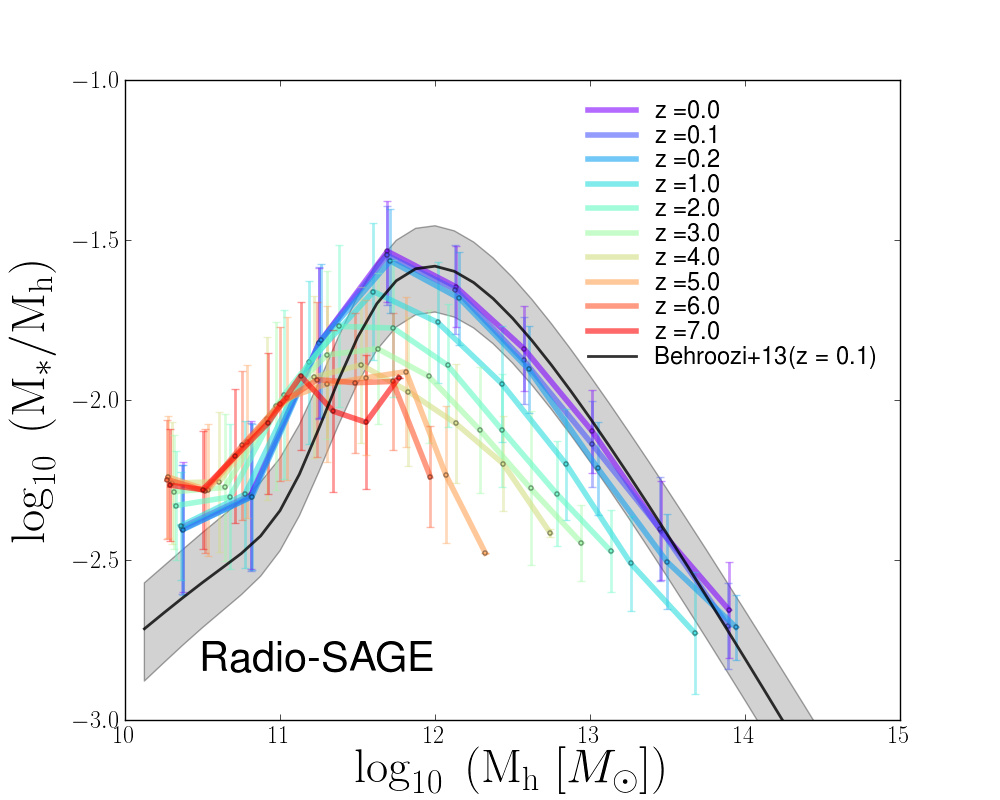}
		\includegraphics[width=0.45\textwidth]{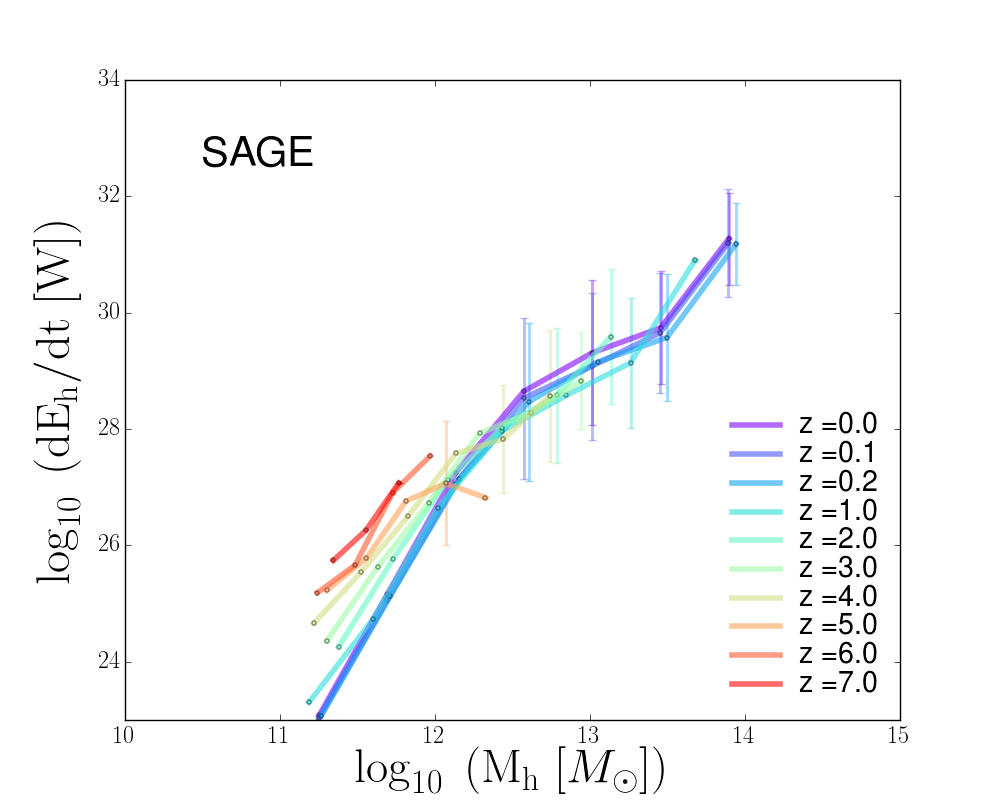}
		\includegraphics[width=0.45\textwidth]{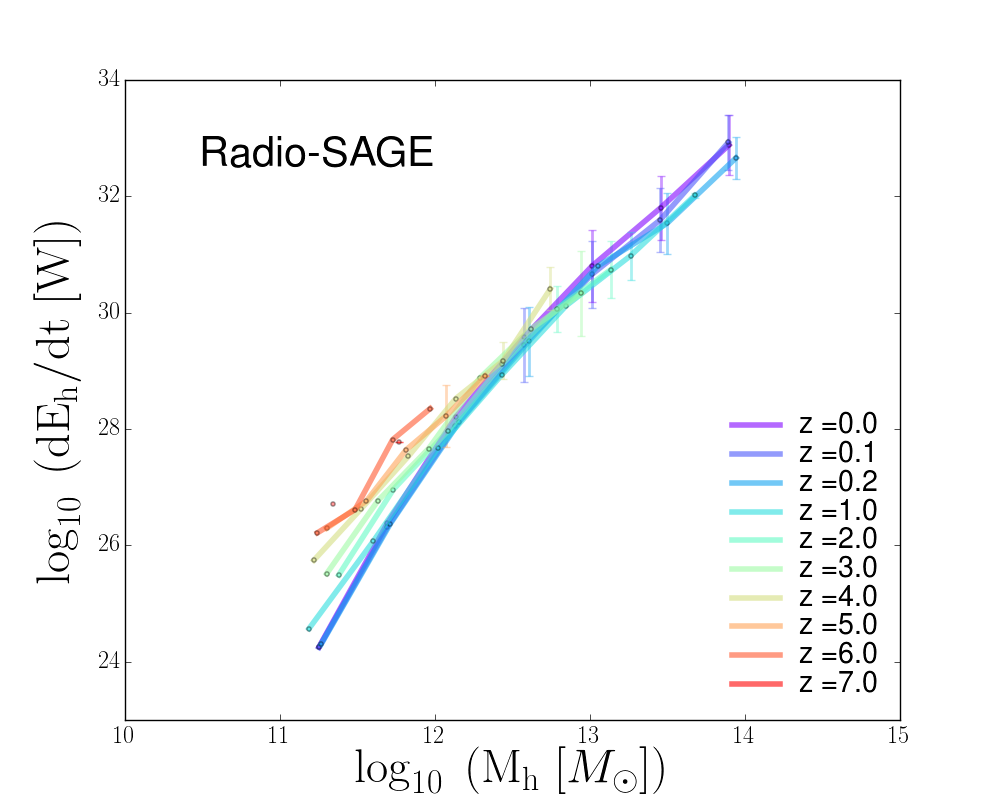}
		\caption{Top: Evolution of the derived stellar mass fractions ($M_*/M_{\rm h}$) as a function of halo mass for  SAGE (left) and \textsc{Radio-SAGE} (right).  Black lines with grey shaded uncertainties are the comparison with \citet{Behroozi2013a,Behroozi2013b} at $z = 0.1$. Bottom: Evolution of the average energy injection rate, ${\rm d}E_{\rm h}/{\rm d}t$,  as a function of halo mass for \textsc{Radio-SAGE} (right) and SAGE (left) in various redshift from  $z = 1$ to $z = 7$. In each case, the lines show the mean values for central galaxies, including statistical uncertainties .}
		\label{fig:Mhalo}
	\end{figure*}     

	\subsection{Feedback process}
	
	Massive dark matter halos can contain substantial amounts of hot gas that should lead to runaway cooling at rates that are unsupported by the observations \citep{Thoul1995,Benson2003}. Our model provides an energy counterbalance to such cooling through the heating resulting from the feedback process. 
	The effectiveness of this feedback, and at the right mass scale, is usually quantified using the galaxy stellar mass function, the shape of which is modified by both AGN and supernovae. The low mass end is constrain by supernovae feedback processes as described in \citet{Croton2006}. The high mass end constrained provided by AGN feedback, a characterised by intermittent black hole accretion and resultant feedback \citep{Shabala2009}.
	
   As fully described in \cite{Raouf2017}, AGN feedback is followed through different phases, which include the turning on of jets to inflate a cocoon due to black hole accretion, cessation of accretion and hence the jet which deflates the cocoon, and finally a quiescence phase of galaxies until black hole accretion begins again. The model allows cooling of gas onto the galaxy only during the quiet phase, with the duration of this time set by our prescriptions for the AGN jet on, off, and cocoon return times.

	Within our model gas heating from an AGN jet acts to raise the mean temperature of the hot halo by up to $\sim 20\%$ above the virial temperature, following the approach of \cite{Shabala2009}. The amount of energy available for feedback is set by the observed properties of radio AGN populations, which are mapped to physical parameters (jet power, active lifetime) using the environment-sensitive formalism of \citet{Shabala2013}. In this model we are using $\sim$ 35\% of accretion power to providing the necessary energy input to quench star formation and reproduce the observed local stellar mass function, the evolution of star formation density, net cooling rate temperature relation, and radio luminosity function all at the same time \citep{Raouf2017}.

\subsection{AGN outflow rate}
When the cocoon expands, some of the jet energy goes to the relativistic cocoon plasma, and some to the surrounding gas (thermal, kinetic and gravitational components). To drive outflows, we are interested in the kinetic component. Simulations \citep[e.g. ][]{Hardcastle2013,Hardcastle2014,Yates2018}  suggest that between 50 and 80 percent of the jet energy (depending on environment, more in low-mass systems such as galaxy groups, less in large clusters) goes to the hot intracluster medium(ICM), and between 10-35 percent of that value \citep[i.e. 5-28 percent of the total jet input energy - see right panel of figure 9 of][]{Hardcastle2013} goes to the kinetic component of the ICM. The feedback power is given by

\begin{equation} 
\label{eq-Edot-Feed} 
Q_{\rm jet}	 =\eta\ \dot{m}_{\rm BH} c^2,
\end{equation} 
where $c$ is the speed of light and $\eta$ is the jet efficiency which we constrain by observations to 0.35 in our model \citep[see; ][]{Raouf2017}. The outflow energy can be written by
	\begin{equation} 
	\label{eq-Edot-Feed} 
	\dot{E}_{\rm w} \equiv \epsilon\ Q_{\rm jet},
	\end{equation} 
	where $\epsilon$ is the outflow coupling efficiency which we are using the upper limit of 0.3 \citep{Hardcastle2013}.
	The outflowing gas is driven outward at a velocity $v_{\rm w}$ and mass outflow rate $\dot{M}_{\rm w}$ \citep{Barai2014, Barai2016}. Given the energy-conservation equation, 
	\begin{equation} 
	\frac{1}{2} \dot{M}_{\rm w} v_{\rm w}^2 = \dot{E}_{\rm w}, 
	\end{equation} 
	the outflow rate can be expressed in terms of the black hole accretion rate, 
	\begin{equation} 
	\label{eq-MdotW-EDW} 
	\dot{M}_{\rm w} ({\rm AGN}) \equiv 2 \frac{\epsilon\ Q_{\rm jet}}{v_{\rm w}^2}\ \left[\frac{M_{\odot}}{\rm yr}\right] =  2 \epsilon \eta\ (\frac{c}{v_{\rm w}})^2\ \dot{m}_{\rm BH}.	
	\end{equation} 
 Here, $v_{\rm w}$ is given by
	\begin{equation}
	v_{\rm w} \equiv \dot{r}_{\rm shock} = {3 \over  5-\beta} {r_{\rm shock} \over t_{\rm on}},
	\end{equation} 
	where $\beta$ describes the variation of inner and outer slopes of the density profile, $t_{\rm on}$ is approximately given by \cite{Turner2015}
	\begin{equation}\label{eq:t_on}
	t_{\rm on} = 120 ~ \left[{m_* \over 10^{11} \rm M_{\odot}}\right]^{0.7} ~ \rm Myr,
	\end{equation} 
	and $r_{\rm shock}$ is the maximum radius of the shocked gas describe in \cite{Raouf2017}.
	The value of $v_w$ is motivated by typical AGN wind velocities seen in observations with 
	a few $1000$ to $10000$ km/s \citep[e.g.,][]{Ramirez2008, Perna2015, Williams2017}. 

Here we arguing about the giving kinetic energy to the hot gas (and driving outflows in that), rather than the cold molecular/atomic gas.
Note that the \citet{Kaiser1997} models effectively assume momentum-conserving outflows. This is the basis for calculating $\dot{r}_{\rm shell}$. The equation for wind outflows uses energy conservation, but it ignores thermal heating (and cooling) in the shock (on the one hand), and work done in adiabatically expanding the cocoon (on the other). \citet{Alexander2002} showed that thermal conductivity in the swept up shell will make a difference to the first point (i.e. whether the gas stays hot or cools rapidly). 
In our model, the wind estimation doesn't play an active role in the modeling; we simply use it to calculate the mass outflow rate.	

\subsection{Star formation rate}
The gas that does manage to cool pools in the galactic disk, and forms stars. The star formation rate can now be calculated from a Kennicutt-Schmidt-type relation \citep{Kennicutt1998}:
\begin{equation}
\dot{m}_* = \alpha_{\rm{SF}} \, \frac{(m_{\rm cold} - m_{\rm crit})}{t_{\rm dyn,\rm disk}}~,
\label{sfr}
\end{equation}
where $m_{\rm cold}$ is the total mass of cold gas and $\alpha_{\rm{SF}}$ is the star formation efficiency $\sim0.05$ in our model. In other words, a fraction $\alpha_{\rm{SF}}$ of gas above the threshold is converted into stars in a disk dynamical time $t_{\rm dyn,{\rm disk}} = r_{\rm disk}/V_{\rm vir}$. In this work, this star formation rate applies to both bulges and disks of galaxies.

	In this paper, we focus on observed global galaxy properties and correlations, centred around star formation. Our observable figures are the star formation rate (SFR) function, the evolution of specific star formation rate (sSFR) and the star forming main sequence, shown and discussed below in Section \ref{Sec:Constrains}. We compare our results to a model without AGN to show the importance of feedback at different cosmic epochs and stellar masses. We also show the improvement of our model in comparison to the original SAGE in the evolution of the stellar mass --halo mass relation. All model results assume a Universe where ${h = 0.73}$, and when relevant, a Chabrier initial mass function \citep{Chabrier2003}  to compare the model to the observations.

	\section{Model constraints}\label{Sec:Constrains}
	To calibrate this model we use a set of observables that the output must reasonably compare with, and in a physically sensible way. Only then can we extend our analysis to explore predictions and consequences that can be compared with future observations. Note that the standard approach of matching \textsc{Radio-SAGE} with observations at $z=0$ was presented in our previous paper \citep{Raouf2017} as the calibration of our model, and here we just focus on the model SFRs in the redshift range $z = 0$ to $z = 7$.

	\subsection{Evolution of the star formation rate }
	
	Figure ~\ref{fig:SFRF} shows the SFR function of \textsc{Radio-SAGE} model for the redshift ranges of $z = 0-1$.  In the above figure the curves with shaded regions are the observations by \citet{Gruppioni2015} for same redshift range. 
	The SFR function is only reasonably well modelled to $ z\sim0.5$, while the model is too efficient at quenching star formation at $z=0.5-1$. Appendix Table \ref{tab:SFRF}1 reports the results from the above figure. 
	
	The redshift evolution of the median specific star formation rates is shown in Figure \ref{fig:SFR_Z}  for different stellar mass bins between $\log_{10}(M_{*} /[M_{\odot}]) =$ 8 and 11.5. The trends of \textsc{Radio-SAGE} sSFR in different stellar mass bins are consistent with observed data points at each redshift, in contrast to the model without AGN feedback. In our model, AGN feedback affects galaxies of all masses, albeit at different efficiencies. The results are tabulated in Appendix Table \ref{tab:SFR_Z}2.
	
	The global evolution of the correlation between the sSFR and stellar mass, the sSFR main sequence, shows in Figure \ref{fig:SFR_MS}. The left panel of Figure \ref{fig:SFR_MS} shows the sSFR sequence for different redshift bins in the range $z = 0-6$.  We compare our model with observations of \cite{Elbaz2007} for redshift range between 0.0 and 0.1, \cite{Oliver2010} for redshift range between 0.1 and 0.2 and  \cite{Pearson2018} for redshift down to 6 (colour dashed-lines). 
	The evolution of the sSFR sequence in our model is a close match to the observational trend. Further, to address in detail the question of how AGN feedback affects the star-formation main sequence, we make a similar plot for the no-AGN feedback model in the right panel of Figure \ref{fig:SFR_MS}. Figure  \ref{fig:SFR_MS} shows that radio jets are important at both low and high (e.g. $z>$3) redshift, where the two models predict different trends in specific star-formation main sequence.  In Figure \ref{fig:AGN_Scale} we also show the median stellar mass at each redshift, for two different narrow final halo masses for the models with and without AGN;
	this plot illustrates the mass scales on which AGN feedback becomes important at each redshift.
	
	\subsection{ Comparison with original SAGE}
	
	As described in section \ref{ModelDescribtion}, the main differences of \textsc{Radio-SAGE} with the original \textsc{SAGE} model are in a more sophisticated density profile and intermittent feedback from AGN (by switching off the ``radio mode" of the original \textsc{SAGE} model of \cite{Croton2006}). In light of these differences, it is important to consider the consistency of model predictions with recent observations. In comparison to the original \textsc{SAGE}, our model has better agreement with the studies by \cite{Behroozi2013a} of the relationship between stellar mass and halo mass. Sub-panels of Figure \ref{fig:Mhalo} show comparison of both \textsc{Radio-SAGE} and \textsc{SAGE} model for such a relationship. The top panels present the evolution of the derived stellar mass fractions ($M_*/M_{\rm h}$) as a function of halo mass (parameters for each model are reported in Appendix Tables \ref{tab:MHMSMH1}5, \ref{tab:MHMSMH2}6) . In all panels, the black lines with grey shaded error regions are the comparison with \cite{Behroozi2013a,Behroozi2013b}. As can be seen, the trend of \textsc{Radio-SAGE} have more consistency with  \cite{Behroozi2013a} at the present epoch, and follows the observationally comparable trend of higher redshift evolution \citep[see Fig. 7 in ][]{Behroozi2013a} in comparison to the original \textsc{SAGE}. The bottom panels of the Figure \ref{fig:Mhalo} show the evolution of the average energy injection rate as a function of halo mass. We calculate this energy, $\rm dE_h/dt$, by multiplying the energy injected per outburst, $Q_{\rm jet}$, by the AGN duty cycle, $t_{\rm on} \delta$ where $\delta= 0.05 \left[{m_* / 10^{11}\rm M_{\odot}}\right]^{1.5}$ \citep{Best2005}, then divide by the time between redshift slices \citep[For definition of these quantities see; ][]{Raouf2017}. The figure shows \textit{where} the energy is being injected, eventually resulting in different stellar masses in the two models: in \textsc{Radio-SAGE}, more energy is deposited in massive ($>10^{13} M_\odot$) haloes since $z = 2$. Hence, our model is applying feedback in the ``right" galaxies due to the mass-dependence of the radio AGN trigger compared to the original \textsc{SAGE} model.
	\begin{figure*}
	\centering
	\includegraphics[width=1.0\textwidth]{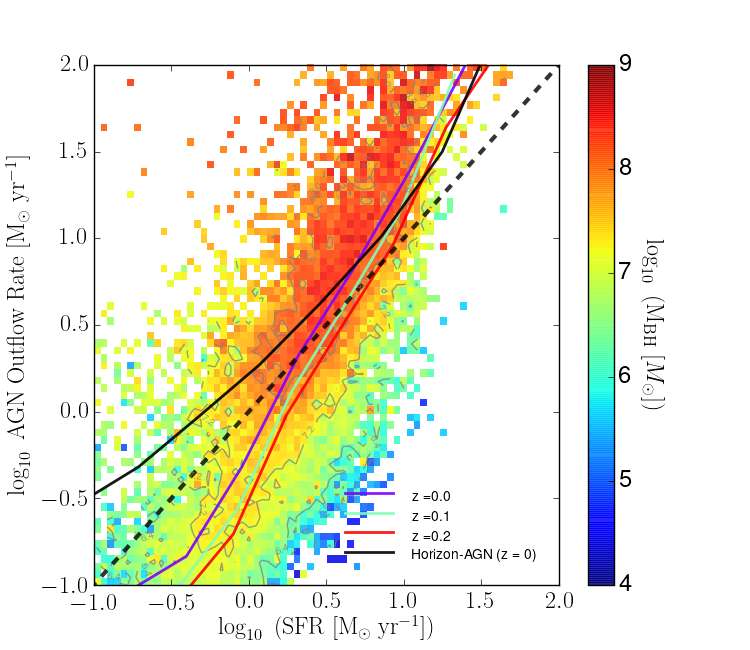}
	\caption{ The local ($z<0.2$) outflow rate versus the SFR  of galaxies, color coded by black hole mass. Coloured lines are medians in three redshift bins. The dashed line show  1:1 correspondence between the SFR and outflow rate. The black solid curve shows the median relation from Horizon-AGN hydrodynamical simulation . 
	}
	\label{fig:Outflow}
\end{figure*} 

\begin{figure*}
	\centering
	\includegraphics[width=.95\textwidth]{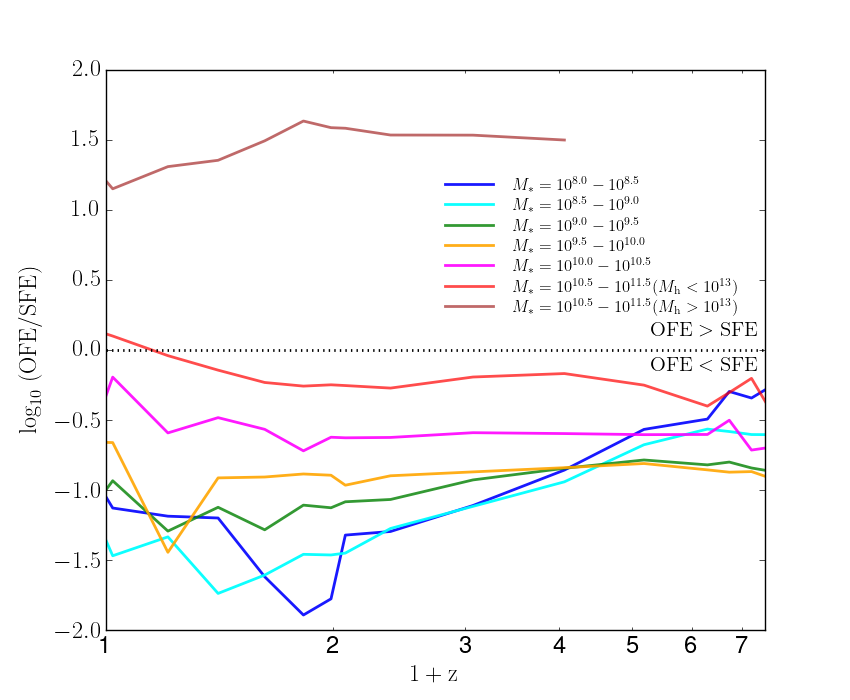}
	\caption{ The redshift evolution of mass loading factor, equivalent to the ratio of outflow efficiency (OFE) to the star formation efficiency (SFE) of galaxies. The dashed line shows 1:1 correspondence between the OFE and SFE . 
	}
	\label{fig:SFE}
\end{figure*} 
	
	\section{Results}
	\subsection{Mass loading during the SFR duty cycle}
	\label{Sec:Massloading }
	
	The gas content of the galaxy could be influenced directly by AGN feedback which expels the interstellar medium (ISM) out of galaxies in massive galactic winds, and/or prevents star formation by directly heating the ISM gas \citep{Springel2005,DiMatteo2005,Murray2005,Fabian2006}. 
	While we are using a fraction of accretion power to quench star formation, massive galaxies have significant outflows depend on their black hole mass. Observational evidence supports the notion of frequent and fast outflows in massive galaxies \citep{Tremonti2007} which remove large amounts of gas \citep{Heckman2000,Veilleux2005,Weiner2009,Sturm2011} by using low fraction of accretion power \citep{Moe2009,Saez2009,Dunn2010}.  Knowledge of the rate at which gas is blown out during the feedback process allows us to compute and analyse the \emph{mass-loading factor} (outflow rate / SFR); the rate at which cold gas is removed from galaxies due to AGN activity.
	
	Figure \ref{fig:Outflow} shows the outflow rate as a function of SFR of central galaxies with stellar mass over $10^9 \rm M_{\odot}$, color coded by black hole mass for low-redshift galaxies. 
    There is a correlation between outflow and star formation rate across the mass scales. At lower black hole masses, the SFR is higher than the outflow rate; a similar result has been found in the HORIZON-AGN simulation \citep{Dubois2014} that is not exclusively due to AGN feedback. Note that outflow rates in Horizon-AGN are calculated by summing over cells contained within a narrow shell of a given radius, centred on the halo. The outflow rate per cell is calculated as $\rm \dot{\rm M}_{\rm gas}=\sum_i \rho \Delta x_i^3 \bar v_i \cdot \bar r_i/\omega$, where  $\rho$ is the gas density, $\rm \Delta x$ is the cell size, $\rm \bar v_i$ is the gas velocity, $ \rm \bar r_i$ is the unit vector of the cell centre relative to the halo centre and $\rm \omega = 2kpc $ is the width of the shell. The total outflow rate for a given halo is found by summing $\rm \dot{M}_{gas}$ for all cells within the shell that have $\rm \bar v_i \cdot \bar r_i > 0$. The model has good agreement with the HORIZON-AGN outflow rates at the high SFR end (i.e. the high mass end, where AGN driven outflows are dominant). At lower SFRs (and lower masses), the model for AGN winds falls short compared to the hydrodynamical simulation that includes winds from stellar feedback.
    
	In Figure  \ref{fig:SFE}, we examine the cosmological evolution of the star-formation efficiency, SFE=SFR/M$_{\rm gas}$, and outflow efficiency, OFE=Outflow rate /M$_{\rm gas}$. 	

	In general, at a given redshift, OFE/SFE increases with stellar mass. In the galaxies with mass range of $M_{*} < 10^{10.5} M_{\odot}$, the SFE consistently exceeds OFE. The most massive galaxies in the low mass clusters ($10^{10.5} \leq M_{*} < 10^{11.5}  M_{\odot} ,\ M_{\rm h} < 10^{13}M_{\odot}$), the OFE approximately equals to SFE; while in the massive clusters ($10^{10.5} \leq M_{*} < 10^{11.5}  M_{\odot},\ M_{\rm h} > 10^{13}M_{\odot}$) have higher OFE than SFEs. There is very little obvious redshift evolution in these strong mass trends. Splitting the most massive galaxies by environment (through halo mass) shows that the feedback is prevalent in the most massive galaxies clusters ($M_{\rm h} > 10^{13} M_{\odot}$), precisely the locations where the cooling catastrophe must be prevented. Note that we show the evolution of low mass systems in the above figure just for comparison, although  galaxies with stellar mass less than $10^9 \rm M_{\odot}$ are at the resolution limit of this study.
	
	Further, Figure A1 (appendix) shows the correlation between galaxies mass loading factor, OFE/ SFE, as a function of the black hole mass color coded by halo mass.  We also show the trend of the residuals of the best fit for the OFE/SFE -- $M_{\rm BH}$ relation in different gas fraction ($F_{\rm gas}$), stellar mass and halo mass in the bottom panels of the above figure showing more sensitivity to the halo mass, especially in massive halos, with respect to $F_{\rm gas}$ and $\rm M_*$. Consequently, Figure A1 clearly shows that the ratio of outflow rate to SFR is mainly driven by black hole mass and virial halo mass, $M_{\rm vir}$. 
			
	\section{Discussion and Conclusions}
	
	In this work, we have presented the effect of black hole growth and feedback on star formation properties of galaxies using the \textsc{Radio-SAGE} model. The model implements AGN feedback more realistically and self-consistently in the cooling-heating cycle which affect the process of star formation. We show a close fit to a number of key galaxy population statistics including the star formation rate function, the evolution of specific star formation rate and the specific star formation rate sequence. Expanding in the number of observables for our model, able us to make predictions for radio AGN and galaxies to be used in large-scale observing programs.  
	
	Our results in this  paper are as follows. Our model is consistent with the star formation rate function up to $z = 1$ and makes reasonable predictions for galaxy quenching in this redshift range. 
	At higher redshifts, AGN feedback limits galaxy SFRs to lower rates than the model without AGN (Figure \ref{fig:SFRF}). 
	
	We find that the trend of specific star formation rate evolution with redshift is consistent with observations (Figure \ref{fig:SFR_Z}). The AGN heating  impacts the SFR over all redshift ranges of galaxies between $10^{8}$ to $10^{11.5} M_{\odot}$, and at all redshifts. 
		
   \textsc{Radio-SAGE} shows better agreement with observations of the stellar mass - halo mass relation than the original SAGE model (Figure \ref{fig:Mhalo}). Due to the mass-dependent radio AGN fraction trigger, the \textsc{Radio-SAGE} model is applying  feedback in the ``right" galaxies, while SAGE is doing too much feedback in massive galaxies but not enough in low-mass ones.  We suggest this could be due to the different trends of energy injection in various halo mass for the \textsc{SAGE} model (bottom panels of Figure \ref{fig:Mhalo}). 
	
	We calculate the outflow rate as a function of the star formation rate in various black hole mass bins, and find these to be broadly consistent with cosmological hydrodynamic simulations. The star formation rate is higher than the outflow rate for galaxies with a low black hole mass. We also show the quantity of star formation and outflow efficiency in different redshift and stellar mass bins of galaxies. In general, the star formation efficiency is higher than the outflow efficiency, except for the massive galaxies which host the most massive black holes and are located in massive halos, rapidly cooling environments.

	Super-massive black hole masses are strongly correlated with their host bulge stellar mass and the amount of cold gas present in a merging system plays a large part in how rapidly the black hole and bulge can grow. Future studies will focus the evolution of the $m_{\rm BH}\,$--$\,m_{\rm bulge}$ relation.

	\section*{Acknowledgements}
	We thank the referee for constructive comments and suggestions which helped to improve the paper. MR is grateful to Balzan center for cosmological studies award and the Institut d'Astrophysique de Paris for hosting his stay during which much of this work was carried out. The Semi-Analytic Galaxy Evolution (SAGE) model that served as a basis for this work is a publicly available codebase that runs on the dark matter halo trees of a cosmological N-body simulation. It is available for download at https://github.com/darrencroton/sage.
	The Millennium Simulation on which the semi-analytic model was run was carried out by the Virgo Supercomputing Consortium at the Computing Centre of the Max Plank Society in Garching. It is publicly available online at http://www.mpa-garching.mpg.de/Millennium/ through the German Astrophysical Virtual Observatory.
	The Horizon project was supported by grant  ANR-13-BS05-0005 of the French Agence Nationale de la Recherche and grant NSF PHY11- 25915 by the National Science Foundation.

\appendix
\section{Extra Figures and Tables}
\label{app}

\begin{table*}
	\label{tab:SFRF}		
	\caption{Parameters of Figures \ref{fig:SFRF}, with units of $M_{\odot}/ \rm yr$ and $\rm Mpc^{-3}\ dex^{-1}$ for SFR and $\phi$, respectively.}
	\begin{tabular}{lcccccc}
		\hline \\	
		$\log_{10}({\rm SFR})$     & $\phi(z = 0)$  & $\phi(z = 0.2)$& $\phi(z = 0.4)$&$ \phi(z = 0.6)$& $\phi(z = 0.8)$& $\phi(z = 1.0)$ \\
		\hline \hline
		
		--1.01       &  1.29e--02 &  1.60e--02 &  1.75e--02 &  2.07e--02 &  2.28e--02 &  2.26e--02   \\
		--9.11e--01 &  1.16e--02 &  1.51e--02 &  1.77e--02  &  2.12e--02 &  2.29e--02 &  2.40e--02   \\
		--8.09e--01 &  1.06e--02 &  1.35e--02 &  1.66e--02 &  1.93e--02 &  2.21e--02 &  2.41e--02   \\
		--7.07e--01 &  1.00e--02 &  1.26e--02 &  1.51e--02 &  1.87e--02 &  2.01e--02 &  2.33e--02   \\
		--6.05e--01 &  8.84e--03 &  1.17e--02 &  1.38e--02 &  1.69e--02 &  1.99e--02 &  2.10e--02   \\
		--5.03e--01 &  9.25e--03 &  1.05e--02 &  1.12e--02 &  1.56e--02 &  1.74e--02 &  1.90e--02   \\
		--4.01e--01 &  7.88e--03 &  1.02e--02 &  1.14e--02 &  1.42e--02 &  1.60e--02 &  1.70e--02   \\
		--2.98e--01 &  7.66e--03 &  8.66e--03 &  1.12e--02 &  1.24e--02 &  1.40e--02 &  1.67e--02   \\
		--1.96e--01 &  7.61e--03 &  8.66e--03 &  1.04e--02 &  1.12e--02 &  1.28e--02 &  1.43e--02   \\
		--9.46e--02 &  8.15e--03 &  8.70e--03 &  9.48e--03 &  1.05e--02 &  1.21e--02 &  1.31e--02   \\
		7.44e--03 &  7.68e--03 &  8.11e--03 &  9.13e--03 &  1.06e--02 &  1.13e--02 &  1.21e--02   \\
		1.09e--01 &  6.94e--03 &  7.56e--03 &  8.90e--03 &  9.64e--03 &  1.11e--02 &  1.14e--02   \\
		2.11e--01 &  6.64e--03 &  7.93e--03 &  8.71e--03 &  9.22e--03 &  9.67e--03 &  1.12e--02   \\
		3.13e--01 &  6.10e--03 &  7.23e--03 &  8.65e--03 &  9.54e--03 &  1.04e--02 &  1.07e--02   \\
		4.15e--01 &  5.32e--03 &  7.02e--03 &  7.96e--03 &  9.25e--03 &  9.73e--03 &  9.67e--03   \\
		5.18e--01 &  5.25e--03 &  6.40e--03 &  7.26e--03 &  8.55e--03 &  9.73e--03 &  9.97e--03   \\
		6.20e--01 &  3.93e--03 &  5.38e--03 &  6.89e--03 &  7.99e--03 &  8.33e--03 &  9.83e--03   \\
		7.22e--01 &  3.18e--03 &  4.30e--03 &  5.59e--03 &  6.80e--03 &  8.33e--03 &  8.97e--03   \\
		8.24e--01 &  2.37e--03 &  3.44e--03 &  5.06e--03 &  5.75e--03 &  7.26e--03 &  8.12e--03   \\
		9.26e--01 &  2.07e--03 &  2.93e--03 &  3.58e--03 &  5.22e--03 &  5.60e--03 &  6.67e--03   \\
		1.02        &  1.38e--03 &  2.10e--03 &  3.20e--03 &  3.85e--03 &  4.89e--03 &  5.83e--03   \\
		1.13        &  3.82e--04 &  1.08e--03 &  1.78e--03 &  2.74e--03 &  3.05e--03 &  4.52e--03   \\
		1.23        &  3.18e--04 &  3.34e--04 &  6.53e--04 &  1.40e--03 &  2.61e--03 &  3.04e--03   \\
		1.33      &  1.75e--04 &  2.70e--04 &  4.78e--04 &  7.96e--04 &  1.29e--03 &  1.68e--03   \\
		1.43      &  1.27e--04 &  1.59e--04 &  9.56e--05 &  2.86e--04 &  5.41e--04 &  8.28e--04   \\
		1.53      &  1.11e--04 &  1.11e--04 &  9.56e--05 &  1.75e--04 &  2.70e--04 &  3.18e--04   \\
		1.64      &  1.11e--04 &  7.96e--05 &  3.18e--05 &  1.59e--05 &  9.56e--05 &  1.27e--04   \\
		1.74      &  7.96e--05 &  6.37e--05 &  0.0    &  0.0    &  3.18e--05 &  1.27e--04   \\

		\hline \\
	\end{tabular}
\end{table*}

\begin{table*}
	\label{tab:SFR_Z}		
	\caption{Parameters of Figures \ref{fig:SFR_Z}, with unit of $1/ \rm yr$  for sSFR.}
	\begin{tabular}{cccccccc}
		\hline \\	
		$z$     & $\log_{10}(\rm sSFR1)$  &$ \log_{10}(\rm sSFR2)$& $\log_{10}(\rm sSFR3)$& $\log_{10}(\rm sSFR4)$& $\log_{10}(\rm sSFR5)$& $\log_{10}(\rm sSFR6)$\\ 
		--    &  ($10^{8}$--$10^{8.5}$ $M_{\odot}$) & ($10^{8.5}$--$10^9$ $M_{\odot}$)& ($10^{9}$--$10^{9.5}$ $M_{\odot}$)& ($10^{9.5}$--$10^{10}$ $M_{\odot}$) & ($10^{10}$--$10^{10.5}$ $M_{\odot}$) & ($10^{10.5}$--$10^{11.5}$ $M_{\odot}$)\\ \hline \hline			
		0.   &   --9.18 &   --9.51 &   --9.83 &   --10.00 &   --10.09  &   --10.12  \\
		0.02 &   --9.16 &   --9.51 &   --9.84 &   --9.98  &   --10.07  &   --10.11  \\
		0.20 &   --9.16 &   --9.49 &   --9.75 &   --9.86  &   --9.94   &   --10.01  \\
		0.40 &   --9.11 &   --9.43 &   --9.65 &   --9.72  &   --9.82   &   --9.91  \\
		0.62 &   --9.07 &   --9.38 &   --9.56 &   --9.59  &   --9.70   &   --9.80  \\
		0.82 &   --9.04 &   --9.31 &   --9.46 &   --9.50  &   --9.58   &   --9.69  \\
		0.98 &   --9.01 &   --9.27 &   --9.38 &   --9.40  &   --9.50   &   --9.61  \\
		1.07 &   --8.99 &   --9.24 &   --9.34 &   --9.36  &   --9.47   &   --9.57  \\
		1.38 &   --8.92 &   --9.14 &   --9.21 &   --9.23  &   --9.34   &   --9.47  \\
		2.07 &   --8.75 &   --8.93 &   --8.98 &   --8.99  &   --9.11   &   --9.25  \\
		3.06 &   --8.50 &   --8.66 &   --8.71 &   --8.75  &   --8.88   &   --9.07  \\
		4.17 &   --8.25 &   --8.42 &   --8.47 &   --8.54  &   --8.66   &   --8.73  \\
		5.28 &   --8.03 &   --8.22 &   --8.28 &   --8.37  &   --8.48   &   --8.55  \\
		5.72 &   --7.96 &   --8.14 &   --8.22 &   --8.29  &   --8.37       &     --     \\
		6.19 &   --7.90 &   --8.07 &   --8.17 &   --8.21  &   --8.48        &    --     \\
		6.71 &   --7.83 &   --8.01 &   --8.09 &   --8.16  &   --8.22         &   --     \\
		7.27 &   --7.78 &   --7.95 &   --8.03 &   --8.06  &   --8.54         &   --    \\	
		\hline \\
	\end{tabular}
\end{table*}

\begin{table}
	\label{tab:MHMS1}		
	\caption{Parameters of Figures \ref{fig:Mhalo} the stellar mass as a function of halo mass for $z =$ 0,1,2,3.}
	\begin{tabular}{ccc}
		\hline	
		$\log_{10} (M_{\rm h} /{\rm M}_\odot)$  & $\log_{10} (M_{*}/{\rm M}_\odot)$ & STD error (dex)\\  
		\hline \hline	
		\multicolumn{3}{c}{{$z =$ 0.0}}\\
		\hline
		10.37 & 8.09 & 0.201\\
		10.81 & 8.47 & 0.299\\
		11.25&  9.38 & 0.319\\
		11.69 & 10.10 & 0.207\\
		12.13 & 10.45 & 0.145\\
		12.57 & 10.70 & 0.126\\
		13.01 & 10.86 & 0.123\\
		13.45 & 10.97 & 0.150\\
		13.89 & 11.19 & 0.144\\
		\hline 
		\multicolumn{3}{c}{{$z =$ 1.0}}\\
		\hline 
		10.35 & 8.06 & 0.166\\
		10.77 & 8.44 & 0.297\\
		11.18 & 9.25 & 0.314\\
		11.60 & 9.89 & 0.250\\
		12.01 & 10.23 & 0.199\\
		12.43 & 10.45 & 0.178\\
		12.84 &  10.60 & 0.146\\
		13.26  &10.69 & 0.124\\
		13.67 & 10.90 & 0.057\\
		\hline 
		\multicolumn{3}{c}{{$z =$ 2.0}} \\
		\hline 
		10.32 &  8.05  & 0.161\\
		10.67 &  8.35 & 0.273\\
		11.02 &  9.00  & 0.306\\
		11.37 &  9.57 & 0.284 \\
		11.73 &  9.93  & 0.226\\
		12.08 & 10.16 & 0.197\\
		12.43 & 10.29 & 0.168\\
		12.78 & 10.45 & 0.127\\
		13.13 & 10.58 & 0.128\\
		\hline
		\multicolumn{3}{c}{{$z =$ 3.0}} \\
		\hline
		10.31 &  8.08 & 0.168\\
		10.64 &  8.36 & 0.286\\
		10.97 &  8.92 & 0.296\\
		11.30 &  9.40 & 0.284\\
		11.62 &  9.76 & 0.248\\
		11.95 &  9.99 & 0.204\\
		12.28 & 10.18 & 0.193\\
		12.61 & 10.25 & 0.213\\
		12.94 & 10.42 & 0.196\\
		\hline 
	\end{tabular}
\end{table}

\begin{table}
	\label{tab:MHMS2}		
	\caption{Parameters of Figures  \ref{fig:Mhalo} the stellar mass as a function of halo mass  for $z =$ 4,5,6,7.}
	\begin{tabular}{ccc}
		\hline 
		$\log_{10} (M_{\rm h} /{\rm M}_\odot)$  & $\log_{10} (M_{*}/{\rm M}_\odot)$ & STD error (dex)\\  
		\hline \hline
		\multicolumn{3}{c}{{$z =$ 4.0}} \\
		\hline 
		10.30 &  8.10 & 0.179\\
		10.60 &  8.34 & 0.247\\
		10.91 &  8.82 & 0.287\\
		11.21 &  9.26 & 0.273\\
		11.52 &  9.59 & 0.257\\
		11.82 &  9.80 & 0.247\\
		12.13 & 10.05 & 0.185\\
		12.43 & 10.14 & 0.115\\
		12.74 & 10.24 & 0.075\\
		\hline 
		\multicolumn{3}{c}{{$z =$ 5.0}}\\
		\hline 
		10.27 & 8.08 & 0.181\\
		10.53 & 8.24 & 0.226\\
		10.79 & 8.64 & 0.262\\
		11.04 & 9.03 & 0.271\\
		11.30 & 9.32 & 0.261\\
		11.55 & 9.60 & 0.243\\
		11.81 & 9.87 & 0.228\\
		12.06 & 9.81 & 0.206\\
		12.32 & 9.77 & 0.  \\
		\hline 
		\multicolumn{3}{c}{{$z =$ 6.0}}\\
		\hline 
		10.27 & 8.06 & 0.178\\
		10.51 & 8.22 & 0.209\\
		10.75 & 8.58 & 0.253\\
		10.99 & 8.94 & 0.254\\
		11.24 & 9.26 & 0.255\\
		11.48 & 9.52 & 0.226\\
		11.72 & 9.73 & 0.223\\
		11.96 & 9.73 & 0.122\\
		\hline 
		\multicolumn{3}{c}{{$z =$ 7.0}}\\
		\hline 
		10.29 & 8.08 & 0.162\\
		10.50 & 8.22 & 0.194\\
		10.71 & 8.53 & 0.227\\
		10.92 & 8.81 & 0.229\\
		11.13 & 9.18 & 0.237\\
		11.34 & 9.23 & 0.257\\
		11.55 & 9.38 & 0.224\\
		11.76 & 9.73 & 0.  \\
		\hline 
	\end{tabular}
\end{table}

\begin{table}
	\label{tab:MHMSMH1}		
	\caption{Parameters of Figures \ref{fig:Mhalo} the derived stellar mass fractions ($M_*/M_h$) as a function of halo mass  for $z =$ 0,1,2,3.}
	\begin{tabular}{ccc}
		\hline	
		$\log_{10} (M_{\rm h} /{\rm M}_\odot)$  & $\log_{10} (M_*/M_h)$ & STD error (dex)\\ 
		\hline \hline	
		\multicolumn{3}{c}{{$z =$ 0.0}}\\
		\hline
		10.37 & --2.39 &  0.203 \\
		10.81 & --2.29 &  0.231 \\
		11.25 & --1.82 &  0.235 \\
		11.69 & --1.53 &  0.158 \\
		12.13 & --1.64 &  0.126 \\
		12.57 & --1.83 &  0.132 \\
		13.01 & --2.09 &  0.129 \\
		13.45 & --2.40 &  0.154 \\
		13.89 & --2.65 &  0.149 \\
		\hline 
		\multicolumn{3}{c}{{$z =$ 1.0}}\\
		\hline 
		10.35 & --2.39 &  0.169 \\
		10.77 & --2.29 &  0.229 \\
		11.18 & --1.88 &  0.254 \\
		11.60 & --1.66 &  0.214 \\
		12.01 & --1.75 &  0.189 \\
		12.43 & --1.94 &  0.181 \\
		12.84 & --2.20 &  0.172 \\
		13.26 & --2.51 &  0.148 \\
		13.67 & --2.72 &  0.190 \\
		\hline 
		\multicolumn{3}{c}{{$z =$ 2.0}} \\
		\hline 
		10.32 & --2.33 &  0.169 \\
		10.67 & --2.30 &  0.226 \\
		11.02 & --1.98 &  0.263 \\
		11.37 & --1.76 &  0.253 \\
		11.73 & --1.77 &  0.214 \\
		12.08 & --1.89 &  0.192 \\
		12.43 & --2.09 &  0.176 \\
		12.78 & --2.29 &  0.165 \\
		13.13 & --2.47 &  0.128 \\
		\hline
		\multicolumn{3}{c}{{$z =$ 3.0}} \\
		\hline
		10.31 & --2.28 &  0.177 \\
		10.64 & --2.27 &  0.230 \\
		10.97 & --2.01 &  0.263 \\
		11.30 & --1.85 &  0.263 \\
		11.62 & --1.83 &  0.236 \\
		11.95 & --1.92 &  0.207 \\
		12.28 & --2.09 &  0.196 \\
		12.61 & --2.27 &  0.241 \\
		12.94 & --2.44 &  0.119 \\
		\hline 
	\end{tabular}
\end{table}

\begin{table}
	\label{tab:MHMSMH2}		
	\caption{Parameters of Figures  \ref{fig:Mhalo} the derived stellar mass fractions ($M_*/M_h$) as a function of halo mass  for $z =$ 4,5,6,7.}
	\begin{tabular}{ccc}
		\hline 
		$\log_{10} (M_{\rm h} /{\rm M}_\odot)$  & $\log_{10} (M_*/M_h)$ & STD error (dex)\\ 
		\hline \hline
		\multicolumn{3}{c}{{$z =$ 4.0}} \\
		\hline 
		10.30 & --2.26 &  0.189 \\
		10.60 & --2.25 &  0.218 \\
		10.91 & --2.05 &  0.255 \\
		11.21 & --1.92 &  0.253 \\
		11.52 & --1.88 &  0.242 \\
		11.82 & --1.97 &  0.233 \\
		12.13 & --2.07 &  0.184 \\
		12.43 & --2.19 &  0.149 \\
		12.74 & --2.41 &  0.012 \\
		\hline 
		\multicolumn{3}{c}{{$z =$ 5.0}}\\
		\hline 
		10.27 & --2.24 &  0.191 \\
		10.53 & --2.28 &  0.206 \\
		10.79 & --2.12 &  0.238 \\
		11.04 & --1.99 &  0.256 \\
		11.30 & --1.94 &  0.244 \\
		11.55 & --1.93 &  0.241 \\
		11.81 & --1.91 &  0.234 \\
		12.06 & --2.23 &  0.213 \\
		12.32 & --2.47 &  0.    \\
		\hline 
		\multicolumn{3}{c}{{$z =$ 6.0}}\\
		\hline 
		10.27 & --2.24 &  0.185 \\
		10.51 & --2.28 &  0.192 \\
		10.75 & --2.14 &  0.231 \\
		10.99 & --2.01 &  0.241 \\
		11.24 & --1.93 &  0.241 \\
		11.48 & --1.94 &  0.219 \\
		11.72 & --1.94 &  0.213 \\
		11.96 & --2.23 &  0.158 \\
		\hline 
		\multicolumn{3}{c}{{$z =$ 7.0}}\\
		\hline 
		10.29 & --2.26 &  0.174 \\
		10.50 & --2.28 &  0.184 \\
		10.71 & --2.17 &  0.209 \\
		10.92 & --2.07 &  0.220 \\
		11.13 & --1.92 &  0.231 \\
		11.34 & --2.03 &  0.253 \\
		11.55 & --2.06 &  0.208 \\
		11.76 & --1.93 &  0.    \\
		\hline 
	\end{tabular}
\end{table}

\begin{figure*}
	\centering
	\includegraphics[width=0.75\textwidth]{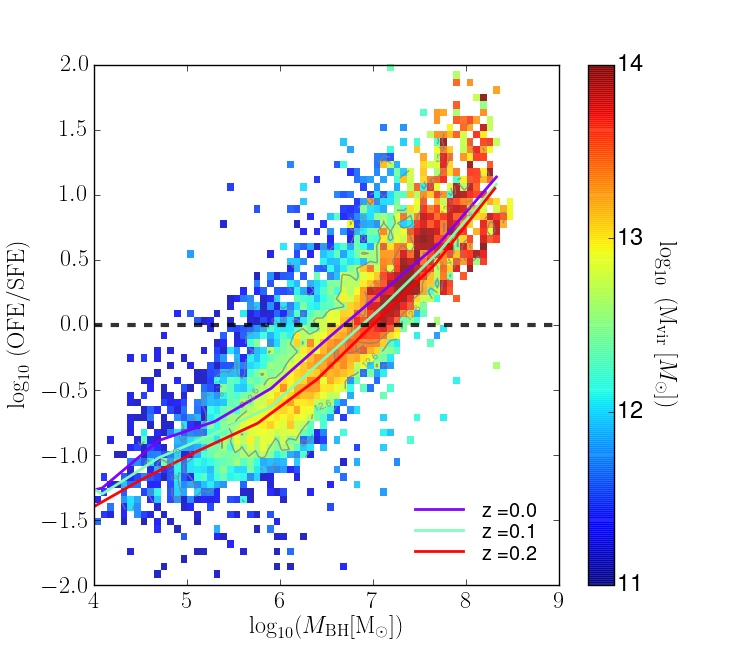} 
	\includegraphics[width=1.1\textwidth]{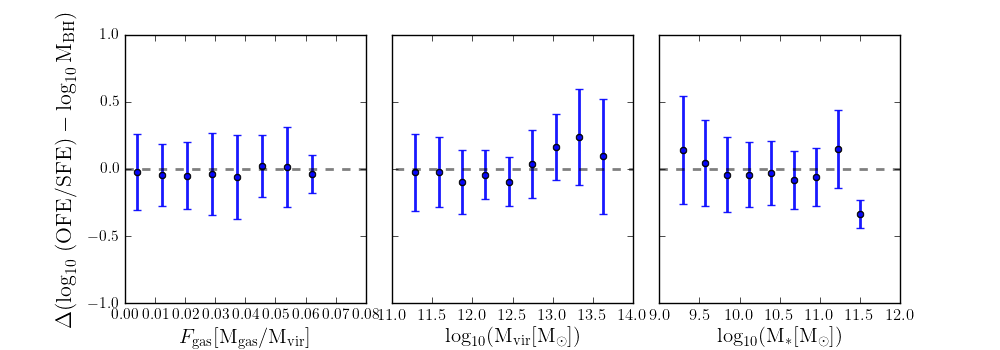}
	\caption{Top: Distribution of galaxies mass loading factor equivalent to the ratio of OFE/SFE as function of black hole mass color code by halo mass, $M_{\rm vir}$. Bottom panels are the residual of OFE/SFE corrected for the best fit relation with $M_{\rm BH}$ as a function of  $M_*$ (right), $M_{\rm vir}$ (middle), the gas fraction $M_{\rm gas}/M_{\rm vir}$ within the virial radius (left) .  }
	\label{fig:Outflow_extra}
\end{figure*}



	\bsp
	\label{lastpage}

\end{document}